\documentclass[12pt]{article}

\usepackage{amssymb}
\usepackage{amsmath}
\usepackage{amsfonts}
\usepackage{amsbsy}
\usepackage{epsfig}
\usepackage{bm,graphicx}

\begin{document}

\begin{titlepage}

\begin{center}

\Huge{Inflationary Hubble Parameter from the Gravitational Wave
Spectrum \\in the General Slow-roll Approximation}

\vskip 1cm

\large{ Minu Joy \footnote{minu@iucaa.ernet.in}}
\\ \vspace{0.5cm}
{\em Inter-University Centre for Astronomy and Astrophysics, Pune,
India }

\vskip 0.5cm

\vskip 1.2cm

\end{center}

\begin{abstract}
 Improved general slow-roll formulae giving the primordial
gravitational wave spectrum are derived in the present work. Also
the first and second order general slow-roll inverse formulae giving
the Hubble parameter $H$ in terms of the gravitational wave spectrum
are derived. Moreover, the general slow-roll consistency condition
relating the scalar and tensor spectra is obtained.
\end{abstract}

\end{titlepage}

\setcounter{page}{0}

\newpage

\setcounter{page}{1}

\section{Introduction}
 Inflation inevitably leads to scalar curvature perturbations
and gravitational waves caused by the tensor perturbations to the
spatial metric \cite{Mukha}. The relative contribution of scalar and
tensor fluctuations to the Cosmic Microwave background (CMB)
anisotropy depends upon the details of the inflationary potential.
Also, the scalar and tensor fluctuations generate different patterns
of polarisation and contribute independently to the observed value
\cite{obs} of the angular power spectrum, $\mathcal{C}_l$. The
exploration of gravitational wave might be possible from the
detection of the B-mode polarisation in the CMB anisotropy \cite{B}
and it is hoped to make more progress in that direction once we get
the more precise data from $\textit{Planck}$ \cite{planck} and
$\textit{Big Bang Observer}$ \cite{BBO}.

The standard slow-roll approximations used for inflationary
scenarios make some strong assumptions about the properties of
inflation, which have not yet been fully confirmed observationally.
Hence, a more general slow-roll approximation has been put forward
\cite{gsr} which lifts the extra, unjustified, assumptions of the
standard slow-roll approximation. The advantages of the general
slow-roll approximation, compared with the standard approximation,
are clearly discussed in Ref.~\cite{ewanrecon}.

The reconstruction of the inflationary potential from the observed
scalar density perturbation spectrum \cite{recon1} and also from the
gravitational wave spectrum have been discussed by many authors
\cite{recon2,recon3}. The nearly scale-invariant spectrum,
$\mathcal{P_\psi}$ of gravitational wave, traces the evolution of
the Hubble parameter during inflation and it has been shown that one
can reconstruct the time dependence of the very early Hubble
parameter and matter energy density from the relic gravitational
wave spectrum \cite{GWrecon}. Recently, we proposed a general
inverse formula \cite{inverse} for extracting inflationary
parameters from the scalar power spectrum. There, we inverted the
single field, general slow-roll formula for the curvature
perturbation spectrum to obtain a formula for inflationary
parameters. Here, our inverse formalism is applied to the
gravitational wave spectrum so as to estimate the Hubble parameter $
H $.

\section{First order general slow-roll formulae}

\subsection{Gravitational wave spectrum}
The formalism of general slow-roll approximation and the power
spectrum calculation are clearly described in \cite{gsr}. With this
general formalism, the first order gravitational wave spectrum can
be given by \cite{JGW}
\begin{equation}\label{FOGWJO}
\ln\mathcal{P_\psi}(\ln k)  =  \int_0^\infty \frac{d\xi}{\xi} \left[
- k\xi \,W'(k\xi) \right] \left[ \ln\frac{1}{p^2} + \frac{2}{3}
\frac{p'}{p} \right]
\end{equation}
where $ p =2\pi a \xi $ and $\xi = - \int \frac{dt}{a} =
\frac{1}{aH} \left( 1 - \frac{\dot{H}}{H^2} + ... \right)$ is minus
the conformal time. The window function $- x \, W'(x)$ is given by,
\begin{equation}\label{W}
W(x) = \frac{3\sin(2x)}{2x^3} - \frac{3\cos(2x)}{x^2} -
\frac{3\sin(2x)}{2x} - 1
\end{equation}
It has the asymptotic behavior
\begin{equation}
\lim_{x \rightarrow 0} W(x) = \frac{2}{5} x^2 + \mathcal{O}(x^4)
\end{equation}
and the window property,
\begin{equation}\label{Wwin}
\int_0^\infty \frac{dx}{x} \left[ - x \, W'(x) \right] = 1
\end{equation}
An alternative form for the gravitational wave spectrum with a
particularly simple window function is
\begin{equation}\label{FOGW1}
\ln\mathcal{P_\psi}(\ln k)  =  \int_0^\infty \frac{d\xi}{\xi} \left[
- k\xi \, v'(k\xi) \right] \left[ \ln\frac{1}{p^2} - 2 \frac{p'}{p}
\right]
\end{equation}
where
\begin{equation}\label{v}
v(x) = \frac{\sin(2x)}{2x} - 1
\end{equation}
$v(x)$ has the asymptotic behavior
\begin{equation}
\lim_{x \rightarrow 0} v(x) = -\frac{2}{3} x^2 + \mathcal{O}(x^4)
\end{equation}
and the window property
\begin{equation}\label{vwin}
\int_0^\infty \frac{dx}{x} \left[ - x \, v'(x) \right] = 1 \,.
\end{equation}
The following figures give the window functions as  the function of
$-\ln (k\xi)$. Large $k\xi$ corresponds to earlier times, when the
mode of interest is within the horizon and oscillates rapidly. The
window function starts to vanish for $k\xi$ $<1$ once the mode
leaves the horizon because it freezes out.
\begin{figure}[!tbh]
   \centering
   \includegraphics[width=10cm,height=6cm,angle=0]{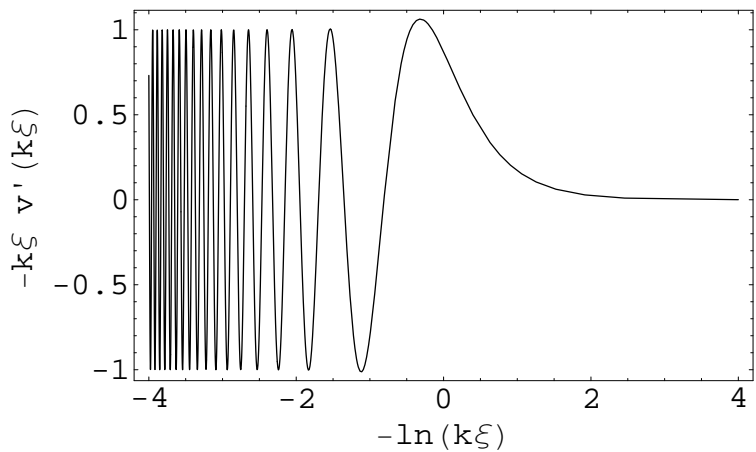}
   \caption{The window function $-k\xi\,v'(k\xi)$ as the function of $-\ln (k\xi)$}
   \label{VP}
\end{figure}
\vspace{1cm}

\begin{figure}[!tbh]
   \centering
   \includegraphics[width=10cm,height=6cm,angle=0]{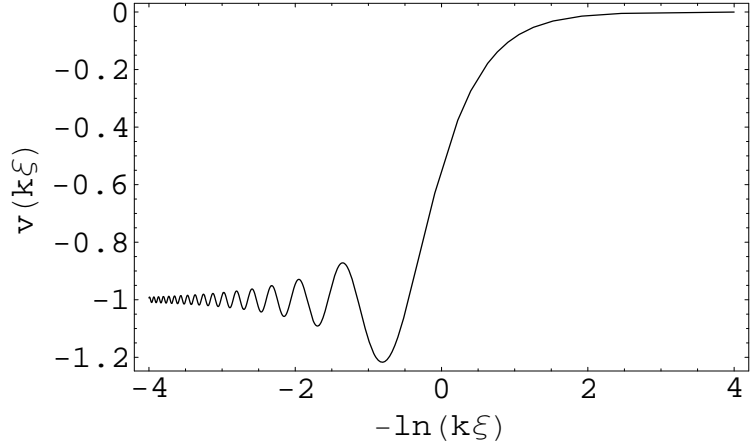}
   \caption{$v(k\xi)$ as the function of $-\ln (k\xi)$}
   \label{VP}
\end{figure}

Here, $p$ is taken as a function of $\ln\xi$ so that
\begin{equation}
p' \equiv \frac{dp}{d\ln\xi} = \xi \frac{dp}{d\xi} =
-\frac{p}{2\pi}\,\frac{dp}{dt}
\end{equation}
and since $ \frac{d\xi}{dt} = -\frac{1}{a} $ and $H =
\frac{\dot{a}}{a} $ we get,
\begin{equation} \label{pH}
\frac{p'}{p} = 1 - \left(\frac{H}{2\pi}\right) p
\end{equation}
Therefore,
\begin{equation}\label{H}
\ln\left(\frac{H}{2\pi}\right)^2 = \ln\left(\frac{1}{p^2}\right) - 2
\left(\frac{p'}{p}\right) - \left(\frac{p'}{p}\right)^2 - \cdots
\end{equation}
Substituting the above form in equation~(\ref{FOGW1}) we get the
general slow-roll formula for gravitational wave spectrum in terms
of the Hubble parameter $H$, up to the first order correction terms,
as
\begin{equation}\label{FOGW}
\ln\mathcal{P_\psi}(\ln k) = \int_0^\infty \frac{d\xi}{\xi} \left[ -
k \, \xi \, v'(k\,\xi)\right] \ln\left(\frac{H}{2\pi}\right)^2
\end{equation}
Also note that $\xi = \frac{1}{a H}$ upto this order.
\subsection{Inverse}
Using the inverse identity,
\begin{equation}\label{id1}
\int_0^\infty \frac{dk}{k} \, m(k\zeta) \, v(k\xi) =
\frac{1}{2\zeta} \left[(\zeta-\xi)
\,\mathrm{sgn}(\zeta-\xi)+(\zeta+\xi)\, \mathrm{sgn}(\zeta+\xi)
\right] - \frac{1}{\zeta} \left[\zeta \,\mathrm{sgn}(\zeta)+ \xi
\,\mathrm{sgn}(\xi)  \right]
\end{equation}
and its derivative with respect to $\xi$,
\begin{equation}\label{id2}
\int_0^\infty \frac{dk}{k} \, m(k\zeta) \left[ - k\xi \, v'(k\xi)
\right] = \frac{\xi}{2\zeta}  \left[ \mathrm{sgn}(\zeta-\xi) -
\mathrm{sgn}(\zeta+\xi) \right]+\frac{\xi}{\zeta}\mathrm{sgn}(\xi)
\end{equation}
where $\mathrm{sgn}(x)=-1$ for $x<0$ and $\mathrm{sgn}(x)=1$ for
$x>0$ and also where
\begin{equation}
m(x) = \frac{2}{\pi} \left[ \frac{1}{x} - \frac{\cos(2x)}{x} -
\sin(2x) \right]
\end{equation}
we get the first order inverse expression for the gravitational wave
spectrum as
\begin{equation}\label{invH}
\ln\left(\frac{H}{2\pi}\right)^2 = \int_0^\infty \frac{dk}{k} \,
m(k\,\xi) \left[\ln\mathcal{P_\psi} -
\frac{\mathcal{P_\psi}'}{\mathcal{P_\psi}} \right] \,.
\end{equation}
It will be interesting to note the asymptotic behaviour
\begin{equation}
\lim_{x \rightarrow 0} m(x) = \frac{4}{3\pi} x^3 + \mathcal{O}(x^5)
\end{equation}
and the window properties
\begin{equation}\label{mwin}
\int_0^\infty \frac{dx}{x} \, m(x) = 1 \, ,
\end{equation}
\begin{equation}\label{mx^2}
\int_0^\infty \frac{dx}{x} \, \frac{1}{x} \, m(x) = \frac{2}{\pi} \,
.
\end{equation}
The equation~(\ref{invH}) gives an explicit formula for $ H $ in
terms of the gravitational wave spectrum $ \mathcal{P_\psi} $.
\section{Second order general slow-roll formulae}
Under the general slow-roll formalism \cite{gsr}, the second order
spectrum for the scalar \cite{JY} and tensor perturbations
\cite{JGW} have been calculated. An improved second order general
slow-roll formula for the gravitational wave spectrum in terms of
the Hubble parameter $H$ is presented in this section.
\subsection{Gravitational wave spectrum}
The general slow-roll gravitational wave spectrum can be derived up
to second order terms as \cite{JGW}
\begin{eqnarray}\label{SOGW1}
\ln\mathcal{P_\psi}(\ln k) & = & \int_0^\infty \frac{d\xi}{\xi}
\left[ - k\xi \, W'(k\xi) \right] \left[ \ln\frac{1}{p^2} +
\frac{2}{3} \frac{p'}{p} \right] + \frac{\pi^2}{2} \left[
\int_0^\infty \frac{d\xi}{\xi} m(k\xi) \frac{p'}{p} \right]^2
\nonumber \\ && \mbox{} - 2\pi \int_0^\infty \frac{d\xi}{\xi}
m(k\xi) \frac{p'}{p} \int_\xi^\infty \frac{d\zeta}{\zeta}
\frac{1}{k\zeta} \frac{p'}{p}
\end{eqnarray}
Using equation~(\ref{pH}) and its derivative, we get
\begin{equation}\label{epsilon}
- \frac{\dot{H}}{H^2} = - \frac{\frac{p''}{p} + \frac{p'}{p} \left(
1 - 2 \frac{p'}{p} \right)}{\left( 1 - \frac{p'}{p} \right)^2}
\end{equation}
Now, we can rewrite the equation~(\ref{SOGW1}) to get the second
order general slow-roll gravitational wave spectrum in terms of $H$
as,
\begin{eqnarray}\label{SOGWH}
\ln\mathcal{P_\psi}(\ln k) & = &\int_0^\infty \frac{d\xi}{\xi}
\left[ - k\xi \, v'(k\xi) \right] \,
\ln\left(\frac{H}{2\pi}\right)^2 + \frac{\pi^2}{2} \left[
\int_0^\infty \frac{d\xi}{\xi} n(k\xi) \frac{\dot{H}}{H^2} \right]^2
\nonumber \\ && \mbox{} + 2\pi \int_0^\infty \frac{d\xi}{\xi}
n(k\xi) \frac{\dot{H}}{H^2} \left\{\frac{1}{k\xi}
\frac{\dot{H}}{H^2} - \int_\xi^\infty \frac{d\zeta}{\zeta}
\frac{1}{k\zeta} \frac{\dot{H}}{H^2} \right\} \nonumber \\
\end{eqnarray}
where
\begin{equation}
n(x) = \frac{1}{\pi} \left[ \frac{1}{x} - \frac{\cos(2x)}{x} \right]
\end{equation}
$n(x)$ has the asymptotic behavior
\begin{equation}
\lim_{x \rightarrow 0} n(x) = - \frac{2}{\pi} x + \mathcal{O}(x^3)
\, ,
\end{equation}
window property
\begin{equation}
\int_0^\infty \frac{dx}{x} \, n(x) = 1 \, ,
\end{equation}
and is related to $m(x)$ by
\begin{equation}
m(x) = n(x) - x \, n'(x) \, .
\end{equation}
\subsection{Inverse}
Substituting first order inverse expression, equation~(\ref{invH}),
into equation~(\ref{SOGWH}) and following the same formalism of
Ref.~\cite{inverse} we get the second order inverse formula for
gravitational wave spectrum as,
\begin{eqnarray}\label{SOGWinv}
\ln\left(\frac{H}{2\pi}\right)^2 &=& \int_0^\infty \frac{dk}{k} \,
m(k\xi) \left[\ln\mathcal{P_\psi}(\ln k) -
\frac{\mathcal{P_\psi}'(\ln k)}{\mathcal{P_\psi}(\ln k)} \right]
\nonumber \\ && \mbox{} - \frac{1}{2\pi^2} \int_0^\infty
\frac{dk}{k} \, m(k\xi) \int_0^\infty \frac{dl}{l}
\ln\left|\frac{k+l}{k-l}\right| \frac{\mathcal{P_\psi}' (\ln
l)}{\mathcal{P_\psi}(\ln l)} \int_0^\infty \frac{dq}{q}
\left[\ln\left|\frac{k+q}{k-q}\right| - \frac{2 q k}{q^2-k^2}
\right] \frac{\mathcal{P_\psi}' (\ln q)}{\mathcal{P_\psi}(\ln q)}
\nonumber
\\ && \mbox{} + \int_0^\infty \frac{dl}{l} \int_0^\infty
\frac{dq}{q} \, N(l\xi,q\xi) \, \frac{\mathcal{P_\psi}'(\ln
l)}{\mathcal{P_\psi}(\ln l)} \frac{\mathcal{P_\psi}'(\ln
q)}{\mathcal{P_\psi}(\ln q)}
\end{eqnarray}
where $ N(x,y) = M_1(x,y) - M_2(x,y) $ with
\begin{equation}\label{M1}
\int_0^\infty \frac{d\zeta}{\zeta} \, m(l\zeta) \, \int_0^\infty
\frac{dk}{k^2} \, m(k\xi) \, m(k\zeta) \, \frac{\sin^2(q\zeta)}
{q\zeta^2} = M_1(l\xi,q\xi)
\end{equation}
\begin{equation}\label{M2}
\int_0^\infty \frac{d\zeta}{\zeta} \, m(l\zeta) \, \int_0^\infty
\frac{dk}{k^2} \, [(-k\xi)m'(k\xi)] \, m(k\zeta) \,
\frac{\sin^2(q\zeta)} {q\zeta^2} = M_2(l\xi,q\xi)
\end{equation}
Carrying out the integrations in the above expression,
\begin{eqnarray}
\frac{M_j(x,y)}{j} & = & \frac{2}{\pi^2 x y} \left[g_j(x) + g_j(y) -
\frac{1}{2} \, g_j(x-y) - \frac{1}{2} \, g_j(x+y) \right]
\end{eqnarray}
for which the index $ j $ takes values 1 and 2.
\begin{equation}
g_j(x) = x \, \mathrm{Si}(2x) + \frac{j}{2}\,\left( \cos(2x) - 1
\right)
\end{equation}
where we denote
\begin{equation}
\mathrm{Si}(x) \equiv \int_0^x \frac{\sin t}{t} \, dt
\end{equation}
$M_j(x,y)$ has the window property
\begin{equation}\label{Mwin}
\int_0^\infty \frac{dx}{x} \int_0^\infty \frac{dy}{y} \,
\frac{M_j(x,y)}{j} = 1
\end{equation}
and the asymptotic behaviour
\begin{equation}
\lim_{x,y \rightarrow 0} \frac{M_j(x,y)}{j} = \frac{4 x y}{3\pi^2}
\left[ 1 + \mathcal{O}\left(x^2+y^2 \right)\right]
\end{equation}
\section{General slow-roll consistency condition}
The standard slow-roll consistency condition relating the scalar and
tensor spectra is detailed in \cite{recon2} for the single field
case and \cite{MisaEwan} discusses the same for the multi-component
scalar field inflation models. This section describes the general
slow-roll generalisation of the constraint on the spectra. \\ \\
The inverse formula for the scalar power spectrum \cite{inverse},
$\mathcal{P}_s$ is given by,
\begin{eqnarray}\label{finv}
\ln\frac{1}{f^2} &  = & \int_0^\infty \frac{dk}{k} m(k\xi)
\ln\mathcal{P}_s (\ln k) \nonumber \\  && \mbox{} - \frac{1}{2\pi^2}
\int_0^\infty \frac{dk}{k} \, m(k\xi) \left[ \int_0^\infty
\frac{dl}{l} \ln\left|\frac{k+l}{k-l}\right|
 \frac{\mathcal{P}_s' (\ln l)}{\mathcal{P}_s(\ln l)} \right]^2
\nonumber \\ && \mbox{} + \int_0^\infty \frac{dl}{l} \int_0^\infty
\frac{dq}{q} \, M_1(l\xi,q\xi) \,
 \frac{\mathcal{P}_s'(\ln l)}{\mathcal{P}_s(\ln l)} \frac{\mathcal{P}_s'(\ln q)}{\mathcal{P}_s(\ln q)}
\end{eqnarray}
where $ f = \frac{2\pi a \xi \dot\phi}{H} $\,. Also, for the tensor
spectrum we have,
\begin{eqnarray}\label{pinv}
\ln\frac{1}{p^2} & = & \int_0^\infty \frac{dk}{k} \, m(k\xi)
\ln\mathcal{P_\psi}(\ln k) \nonumber \\  && \mbox{} -
\frac{1}{2\pi^2} \int_0^\infty \frac{dk}{k} \, m(k\xi) \left[
\int_0^\infty \frac{dl}{l} \ln\left|\frac{k+l}{k-l}\right|
\frac{\mathcal{P_\psi}' (\ln l)}{\mathcal{P_\psi}(\ln l)} \right]^2
\nonumber \\ && \mbox{} + \int_0^\infty \frac{dl}{l} \int_0^\infty
\frac{dq}{q} \, M_1(l\xi,q\xi) \, \frac{\mathcal{P_\psi}'(\ln
l)}{\mathcal{P_\psi}(\ln l)} \frac{\mathcal{P_\psi}'(\ln
q)}{\mathcal{P_\psi}(\ln q)}
\end{eqnarray}
Combining the above two inverse formulae we can write,
\begin{eqnarray}\label{phiH}
\ln \left(\frac{\dot\phi}{H} \right)^2 & = & \int_0^\infty
\frac{dk}{k} \, m(k\xi) \ln
\left(\frac{\mathcal{P_\psi}}{\mathcal{P}_s}\right)\nonumber \\  &&
\mbox{}- \frac{1}{2\pi^2} \int_0^\infty \frac{dk}{k} m(k\xi)
\int_0^\infty \frac{dl}{l} \ln\left|\frac{k+l}{k-l}\right| \left(
\frac{\mathcal{P_\psi}' }{\mathcal{P_\psi}} + \frac{\mathcal{P}_s'
}{\mathcal{P}_s} \right) \int_0^\infty \frac{dq}{q}
\ln\left|\frac{k+q}{k-q}\right| \left( \frac{\mathcal{P_\psi}'
}{\mathcal{P_\psi}} + \frac{\mathcal{P}_s' }{\mathcal{P}_s} \right)
\nonumber
\\ && \mbox{} + \int_0^\infty \frac{dl}{l} \int_0^\infty
\frac{dq}{q} \, M_1(l\xi,q\xi) \, \left\{
\frac{\mathcal{P_\psi}'(\ln l) }{\mathcal{P_\psi}(\ln l)}
\frac{\mathcal{P_\psi}'(\ln q) }{\mathcal{P_\psi}(\ln q)}  -
\frac{\mathcal{P}_s'(\ln l) }{\mathcal{P}_s(\ln l)}
\frac{\mathcal{P}_s'(\ln q) }{\mathcal{P}_s(\ln q)} \right\}
\end{eqnarray}
Defining,
\begin{equation}
m_1(x)= \int\frac{dx}{x}\,m(x) =
\frac{2}{\pi}\left[\frac{\cos(2x)}{x} - \frac{1}{x}
+\mathrm{Si}(2x)\right]
\end{equation}
we can rewrite,
\begin{eqnarray}\label{phiH}
\left(\frac{\dot\phi}{H} \right)^2 & \simeq & \left(
\frac{\mathcal{P_\psi}}{\mathcal{P}_s}\right)_\diamond \left[1 -
\int_0^\infty \frac{dk}{k} (m_1(k\xi) - \theta(k\xi - k_\diamond
\xi)) \left(\frac{\mathcal{P_\psi}' }{\mathcal{P_\psi}} -
\frac{\mathcal{P}_s' }{\mathcal{P}_s} \right) \right. \nonumber \\
&& \left. \mbox{}  \hspace{1.5cm} + \frac{1}{2}\left(\int_0^\infty
\frac{dk}{k} (m_1(k\xi) - \theta(k\xi - k_\diamond \xi))
\left(\frac{\mathcal{P_\psi}' }{\mathcal{P_\psi}} -
\frac{\mathcal{P}_s' }{\mathcal{P}_s}\right)\right)^2  \right. \nonumber \\
&& \left. \mbox{}\hspace{1.5cm} - \frac{1}{2\pi^2} \int_0^\infty
\frac{dk}{k} m(k\xi) \int_0^\infty \frac{dl}{l}
\ln\left|\frac{k+l}{k-l}\right| \left(\frac{\mathcal{P_\psi}'(\ln l)
}{\mathcal{P_\psi}(\ln l)} +
\frac{\mathcal{P}_s'(\ln l) }{\mathcal{P}_s(\ln l)}\right) \right. \nonumber \\
&& \left. \mbox{} \hspace{2.5cm} \times \int_0^\infty \frac{dq}{q}
\ln\left|\frac{k+q}{k-q}\right| \left(\frac{\mathcal{P_\psi}'(\ln q)
}{\mathcal{P_\psi}(\ln q)} +
\frac{\mathcal{P}_s'(\ln q) }{\mathcal{P}_s(\ln q)}\right) \right. \nonumber \\
&& \left. \mbox{}\hspace{1.3cm} + \int_0^\infty \frac{dl}{l}
\int_0^\infty \frac{dq}{q} \, M_1(l\xi,q\xi) \, \left\{
\frac{\mathcal{P_\psi}'(\ln l) }{\mathcal{P_\psi}(\ln l)}
\frac{\mathcal{P_\psi}'(\ln q) }{\mathcal{P_\psi}(\ln q)}  -
\frac{\mathcal{P}_s'(\ln l) }{\mathcal{P}_s(\ln l)}
\frac{\mathcal{P}_s'(\ln q) }{\mathcal{P}_s(\ln q)} \right\} \right]
\nonumber \\
\end{eqnarray}
where $k_\diamond$ is some reference wavenumber. Now, taking the
derivative of equation~(\ref{SOGWinv}) with respect to $\ln \xi $ we
get,
\begin{eqnarray}\label{HdotH}
2\frac{\dot{H}}{H^2} & = & \int_0^\infty \frac{dk}{k} \,
m(k\xi)\left[\frac{\mathcal{P_\psi}'(\ln k) }{\mathcal{P_\psi}(\ln
k)} - \left(\frac{\mathcal{P_\psi}'(\ln k) }{\mathcal{P_\psi}(\ln
k)}\right)'\,\right] \nonumber \\ && \mbox{} - \frac{1}{2\pi^2}
\int_0^\infty \frac{dk}{k}\,(k\xi) m'(k\xi) \int_0^\infty
\frac{dl}{l} \ln\left|\frac{k+l}{k-l}\right|
\frac{\mathcal{P_\psi}'(\ln l) }{\mathcal{P_\psi}(\ln l)}
\int_0^\infty \frac{dq}{q} \left[\ln\left|\frac{k+q}{k-q}\right| +
\frac{2 q k}{k^2 - q^2} \right] \frac{\mathcal{P_\psi}'(\ln q)
}{\mathcal{P_\psi}(\ln q)} \nonumber
\\ && \mbox{}+ \int_0^\infty \frac{dl}{l} \int_0^\infty
\frac{dq}{q} \,[ M'_1(l\xi,q\xi)- M'_2(l\xi,q\xi)] \,
\left(\frac{\mathcal{P_\psi}'(\ln l) }{\mathcal{P_\psi}(\ln
l)}\right)' \left(\frac{\mathcal{P_\psi}'(\ln q)
}{\mathcal{P_\psi}(\ln q)}\right)'
\end{eqnarray}
Using the definition of the  slow-roll parameter $\epsilon \equiv
\frac{1}{2}\left(\frac{\dot\phi}{H} \right)^2 = -
\frac{\dot{H}}{H^2}$, one can equate the equations (\ref{phiH}) and
(\ref{HdotH}) to get the general slow roll constraint on the spectra
as,
\begin{eqnarray}\label{cons}
\left(\frac{\mathcal{P_\psi}}{\mathcal{P}_s}\right)_\diamond &=&
-\int_0^\infty \frac{dk}{k} \,
m(k\xi)\left[\frac{\mathcal{P_\psi}'(\ln k) }{\mathcal{P_\psi}(\ln
k)}- \left(\frac{\mathcal{P_\psi}'(\ln k) }{\mathcal{P_\psi}(\ln
k)}\right)'\,\right] \nonumber \\ && \mbox{} \hspace{1cm}
\times\left\{1 + \int_0^\infty \frac{dl}{l} \, (m_1(k\xi) -
\theta(l\xi - l_\diamond \xi))
\left[\left(\frac{\mathcal{P_\psi}'(\ln l) }{\mathcal{P_\psi}(\ln
l)}\right) - \left(\frac{\mathcal{P}_s'(\ln l) }{\mathcal{P}_s(\ln
l)}\right)\right]\right\} \nonumber \\ && \mbox{}+ \frac{1}{2\pi^2}
\int_0^\infty \frac{dk}{k}\,(k\xi) m'(k\xi) \int_0^\infty
\frac{dl}{l} \ln\left|\frac{k+l}{k-l}\right|
\frac{\mathcal{P_\psi}'(\ln l) }{\mathcal{P_\psi}(\ln l)}\nonumber
\\ && \mbox{}  \hspace{1.3cm} \times \int_0^\infty \frac{dq}{q}
\left[\ln\left|\frac{k+q}{k-q}\right| + \frac{2 q k}{k^2 - q^2}
\right] \frac{\mathcal{P_\psi}'(\ln q) }{\mathcal{P_\psi}(\ln q)}
\nonumber
\\ && \mbox{}- \int_0^\infty \frac{dl}{l} \int_0^\infty
\frac{dq}{q} \,[ M'_1(l\xi,q\xi)- M'_2(l\xi,q\xi)] \,
\left(\frac{\mathcal{P_\psi}'(\ln l) }{\mathcal{P_\psi}(\ln
l)}\right)' \left(\frac{\mathcal{P_\psi}'(\ln q)
}{\mathcal{P_\psi}(\ln q)}\right)'
\end{eqnarray}
\subsection{Single field approximations}
In single field approximation\footnote{The general form of
Eq.(\ref{FOGWJO}) will be identical if we consider the multi-scalar
field inflation model also, but with a different interpretation of
the quantity 'p'. Comparison of the single and multifield general
slow-roll formulae in the case of scalar perturbations is discussed
in \cite{HCL}.}, \, $ - \frac{\dot{H}}{H^2} \simeq constant$ and
thus we can simplify the equation~(\ref{SOGWH}) to get an easy
forward formula as,
\begin{equation}\label{SOGWH1}
\ln\mathcal{P_\psi}(\ln k)  = \int_0^\infty \frac{d\xi}{\xi} \left[
- k\xi \, v'(k\xi) \right] \, \ln\left(\frac{H}{2\pi}\right)^2 +
\frac{\pi^2}{2} \left( \frac{\dot{H}}{H^2} \right)^2
\end{equation}
Now integrating the above equation by parts and then differentiating
with respect to $\ln k$ we get,
\begin{equation}\label{nconst}
\frac{\mathcal{P'_\psi}}{\mathcal{P_\psi}}  = \int_0^\infty
\frac{d\xi}{\xi} \left[(-k\xi)v'(k\xi)\right] \,
\left(2\frac{\dot{H}}{H^2}\right) - \frac{\dot{H}}{H^2}
\end{equation}
Thus,  we can find that the gravitational wave spectral index
$n_\psi \equiv \frac{d \ln\mathcal{P_\psi}}{d \,ln k} =
\frac{\mathcal{P'_\psi}}{\mathcal{P_\psi}} \simeq constant $ for the
single filed approximation. \newline \\
Also from equation~(\ref{SOGWinv}) we get the inverse for single
field approximation as,
\begin{equation}
\ln\left(\frac{H}{2\pi}\right)^2 = \int_0^\infty \frac{dk}{k} \,
m(k\xi) \left[\ln\mathcal{P_\psi} -
\frac{\mathcal{P_\psi}'}{\mathcal{P_\psi}}\, \right]  -
\frac{\pi^2}{8} \left(\frac{\mathcal{P_\psi}'}{\mathcal{P_\psi}}
\right)^2
\end{equation}
Again, since $n_\psi \simeq constant $ and neglecting the terms
containing ${n'_\psi}^2$ also, we can simplify the consistency
condition in equation~(\ref{cons}) to get the single field general
slow roll constraint on the spectra,
\begin{eqnarray}\label{Consiml}
\left(\frac{\mathcal{P_\psi}}{\mathcal{P}_s}\right)_\diamond  &=&
-\int_0^\infty \frac{dk}{k} m(k\xi)\,\left(n_\psi - {n_\psi}'\right)
+ \left[ \frac{\pi}{2} - \frac{5}{2}+ \ln(k_\diamond )\right]
n^2_\psi \nonumber
\\ && \mbox{}  + n_\psi \int_0^\infty \frac{dk}{k}
(m_1(k\xi) - \theta(k\xi - k_\diamond \xi)) (n_s-1)
\end{eqnarray}
\subsection{Multi-field consistency condition}
\label{multi}
For the multi-component scalar field inflation models, since the
scalar power spectrum has the contribution from both parallel and
orthogonal components, $P_S \geq P_{single field}$. Thus the tensor
to scalar power spectra ratio will be smaller than the single field
case and so we can write the inequality upto first order corrections
as,
\begin{equation}\label{Conmulti}
\frac{\mathcal{P_\psi}}{\mathcal{P}_S} < -\int_0^\infty \frac{dk}{k}
m(k\xi) \,\left(n_\psi - {n_\psi}'\right)
\end{equation}
Note that in this case we can not assume $\epsilon \simeq constant $
as it is for the single field.
\section{Standard slow-roll approximation}
\label{secssr}
In the context of the standard approximation for slow-roll, the
gravitational wave spectrum has the form
\begin{equation}\label{gwssr}
\ln\mathcal{P_\psi} = \ln\mathcal{P_\psi}_\diamond +
{n_\psi}_\diamond  \ln \left( \frac{k}{k_\diamond} \right) +
\frac{1}{2} {n_\psi}'_\diamond \ln^2
\left(\frac{k}{k_\diamond}\right) + \cdots
\end{equation}
where $n_\psi \equiv \frac{d \ln\mathcal{P_\psi}}{d ln k}$\,, is the
gravitational wave spectral index and $k_\diamond$ is some reference
wavenumber. Applying our inverse formula given by
equation~(\ref{SOGWinv}), using the window properties given by
equations.~(\ref{mwin}) and~(\ref{Mwin}), and the following results
\begin{equation}
\int_0^\infty \frac{dx}{x} \, m(x) \, \ln(x) = \alpha
\end{equation}
\begin{equation}
\int_0^\infty \frac{dx}{x} \, m(x) \, \ln^2(x) = \alpha^2 +
\frac{\pi^2}{12}
\end{equation}
\begin{equation}
\int_0^\infty \frac{dy}{y} \ln\left|\frac{x+y}{x-y}\right| =
\frac{\pi^2}{2}\, , \ \ \texttt{for}\ \ x > 0
\end{equation}
where $\alpha = 2 - \ln 2 - \gamma \simeq 0.7296$, we get
\begin{equation}\label{ssrinv1}
\ln\left(\frac{H}{2\pi}\right)^2 = \ln\mathcal{P_\psi}_\diamond +
(\alpha_\diamond-1)\,{n_\psi}_\diamond + \frac{1}{2} \left(
\alpha_\diamond^2 -2\alpha_\diamond + \frac{\pi^2}{12} \right)
{n_\psi}'_\diamond + \left( 1 - \frac{\pi^2}{8} \right)
{n_\psi}_\diamond^2 + \cdots
\end{equation}
where $\alpha_\diamond = \alpha - \ln(k_\diamond \xi)$.
Equation~(\ref{ssrinv1}) reproduces the standard slow-roll inverse,
which is trivially obtained from the standard slow-roll formula,
\begin{equation}
\ln\mathcal{P_\psi} = \ln\frac{1}{p_\star^2} - 2 \alpha_\star
\frac{p_\star'}{p_\star} - \left( \alpha^2_\star - \frac{\pi^2}{12}
\right) \frac{p_\star''}{p_\star} + \left(\alpha^2_\star - 4 +
\frac{5\pi^2}{12} \right) \left(\frac{p_\star'}{p_\star}\right)^2 +
\cdots
\end{equation}
where $\alpha_\star = \alpha - \ln(k \xi_\star)$ and $\xi_\star$ is
an arbitrary evaluation point usually taken to be around horizon
crossing. \\ \\
 Now let us deduce the consistency conditions for the standard
slow-roll approximation case. Substituting the standard slow roll
gravitational wave spectrum given by equation~(\ref{gwssr}) in the
general constraint on the spectra given by equation~(\ref{cons}), we
get the constraint, with the standard approximation as,
\begin{eqnarray}\label{conssr}
\left(\frac{\mathcal{P_\psi}}{\mathcal{P}_s}\right) &=& -n_{\psi
*} - (\alpha_* - 1) n'_{\psi *} + (2+\ln k_*) (n_{s
*} -1) n_{\psi *} \nonumber \\ && \mbox{} +
\left(\frac{\pi}{2} - \frac{5}{2} + \ln k_* \right) n^2_{\psi *}
\end{eqnarray}
Comparison of the above expression with the general slow roll
condition shows that the second order correction terms are
different. Also we can write the standard slow roll multi-field
constraint as
\begin{equation}\label{conssr}
\left(\frac{\mathcal{P_\psi}}{\mathcal{P}_S}\right)  < -n_{\psi*}
\end{equation}
It is clear from the general multi-field constraint in equation
~(\ref{Conmulti}) that the sign of second term on right hand side
will depend on that of $ n'_\psi $ and thus the bound could be
varied depending on the tilt of $ n_\psi $.
\subsection*{Acknowledgements}
The author is indebted to Ewan D Stewart for the timely suggestions
and valuable discussions on this work.

\end{document}